\documentclass[final]{Lisbon}
\begin{document}

\title{Chiral symmetry and\\
spectrum of Euclidean Dirac operator}

\author{S\'ebastien Descotes}

\address{Groupe de Physique Th\'eorique, 
I.P.N., 91406 Orsay, France
\\E-mail: descotes@ipno.in2p3.fr}


\maketitle

\abstracts{
After recalling some connections between the Spontaneous Breakdown of Chiral
Symmetry (SB$\chi$S) and the spectrum of the Dirac
operator for Euclidean QCD on a torus, we use this tool to reconsider two
related issues : the Zweig rule violation in the scalar channel
and the dependence of SB$\chi$S order parameters on the number $N_f$
of massless flavours\cite{paramag}. The latter would result into a great
variety of SB$\chi$S patterns in the $(N_f,N_c)$ plane, which could
 be studied through Leutwyler-Smilga sum rules in
association with lattice computations of the Dirac
spectrum\cite{LSSR}.}

\section{Chiral symmetry and its breakdown}

At low energies, QCD cannot be described in a perturbative way : its
degrees of freedom (quarks and gluons) are different from its asymptotic
states (hadrons). To analyse the theory, one can fortunately follow another
path and consider its symmetries. If we set to zero the masses of the
$N_f$ lightest quarks (chiral limit), the right- and left-handed fermions
can be rotated separately in flavour space and QCD exhibits a chiral symmetry
$SU_L(N_f)\times SU_R(N_f)\times U_V(1)$. This symmetry
is spontaneously broken along its axial directions by the
fundamental state of the theory, generating pseudoscalar
Goldstone bosons. When light quark masses are turned on, they appear in the QCD
Lagrangian as small perturbations breaking down explicitly chiral symmetry
and providing the Goldstone bosons a light mass.

In Nature, $N_f=3$ flavours are light enough in comparison with QCD scale
($\sim 1$ GeV) for this treatment, and the Goldstone bosons are
identified with the light pseudoscalar octet $(\pi,K,\eta)$. 
It is often stated that SB$\chi$S is triggered by a large condensation of
quark-antiquark pairs in the vacuum:
$\langle 0|\bar{q}q|0\rangle=-\Sigma(N_f)$,
but this assumption has to (and is about to) be tested
experimentally, at least in the two-flavour case\cite{daphne}.
Supposing the leadership of the quark condensate in SB$\chi$S, the
Standard Chiral Perturbation Theory (S$\chi$PT)\cite{Gasser-Leutwyler}
provides an effective description of the pseudoscalar-octet phenomenology. 
With the same goal, the Generalized $\chi$PT\cite{gchipt}
keeps the quark condensate as a free (and possibly small) parameter and, at 
a fixed order of accuracy, takes into account contributions neglected by
S$\chi$PT.

We highlight the particular status of the two-point correlator
\be
\Xi_{\mu\nu}(q^2)\delta^{ij}=i\int d^4x\ e^{iq\cdot x}
     \langle \Omega | T\{V^i_\mu(x) V^j_\nu(0) - A^i_\mu(x) A^j_\nu(0)\}
     |\Omega\rangle \label{VV-AA}
\ee
where $V$ ($A$) are the Noether currents for vector (axial)
rotations in flavour space, and $|\Omega\rangle$ is the vacuum of
the massive theory. $\Xi_{\mu\nu}$ is a chiral order parameter for any
momentum $q$. In the chiral limit, a massless pole
arises because of the Goldstone bosons coupling to the axial current.
$\Xi_{\mu\nu}(0)$ tends then to $-\eta_{\mu\nu}F^2(N_f)$, yielding
the decay constant of the Goldstone bosons for $N_f$ massless flavours.
A non-vanishing $\Xi_{\mu\nu}(0)$ is
thus a necessary and sufficient condition for SB$\chi$S, by contrast with
other order parameters, whose vanishing is not a clear signal for the full
symmetry restoration.

\section{$N_f$-dependence of SB$\chi$S order parameters}\label{nfdep}

A common approximation leads to neglect the (supposed) weak
$N_f$-dependence of SB$\chi$S order parameters due to light quark loops:
this effect is suppressed in the large-$N_c$ limit and violates the Zweig
rule, both considered as good approximations of QCD\cite{large-Nc}. On the
other hand, the scalar channel does not obey large-$N_c$ predictions, and
all chiral order parameters should vanish for $N_f/N_c$ large enough, where
the perturbative QCD $\beta$-function predicts the end of confinement.
Furthermore, recent lattice simulations\cite{lattflav}
find large variations in the SB$\chi$S pattern when $N_f$ increases, 
and a sum rule estimate\cite{Moussallam},
based on experimental data in the scalar channel, asserts
$\Sigma(N_f)$ could decrease by half from 2 to 3 flavours.

To study this problem, Euclidean QCD on a torus (of volume $V=L^4$)
seems an appealing tool, because SB$\chi$S is related to the lower end of 
the spectrum of the Dirac operator:
$H[G]=\gamma_\mu(\partial_\mu+iG_\mu)$. For each gauge configuration,
this Hermitean operator admits only real eigenvalues (e.v.s) $\lambda_n$.
The spectrum is symmetrical with respect to zero, because
$\{H,\gamma_5\}=0$, and 
the winding number $\nu[G]$ of 
the gauge configuration provides the number of exactly zero e.v.s. We label 
positive e.v.s in the ascending order with a positive
integer, and denote the negative part of the spectrum
$\lambda_{-n}=-\lambda_n$. 

In a
$d$-dimensional space, Dirac e.v.s turn out to be uniformly bound by the free
theory\cite{Vafa-Witten}:
$|\lambda_{n}[G]| < \omega_n \equiv C\cdot n^{1/d}/L$,
where $C$ depends only on the geometry of the
space-time manifold, but neither on $G$, $n$ nor $V=L^d$. This bound shows the
paramagnetic response of the Dirac spectrum to an external gauge field.
We see the lowest e.v.s accumulate around zero for $L\to\infty$.

The eigenvalues $\lambda$ and the eigenvectors $\phi(x)$ of the Dirac operator
are independent of
quark flavours and quark masses, and they can be used to project and integrate
fermionic fields in any functional integral. The following 
propagators and the (normalized) determinants are involved:
\be
S_{j}(x,y|G) =\sum_{n}\frac{\phi_{n}(x)\phi_{n}^{\dagger}(y)}
{m_j - i\lambda_{n}[G]}, \qquad
\Delta(m|G) =m^{|\nu|} \prod _{n > 0}\frac{m^2 + \lambda_{n}^2[G]}{m^2 +
\omega_{n}^2},
\ee
In particular, the fermionic determinant reads:
$\det(M-iH)=\prod_{j}\Delta(m_j|G)$.
In the chiral limit $m_1=\ldots =m_{N_f}=m\to 0$ (the other quark masses
remaining fixed), $\langle\bar{q}q\rangle$ yields\cite{Banks-Casher}:
\be
\Sigma(N_f) = \underline{\lim} \frac{1}{L^4} \ll\!\int\!\!dx\
\mathrm{Tr}\ S_1(x,x|G)\!\gg_{N_f} = 
\underline{\lim} \frac{1}{L^4} \ll\! \sum_{n} \frac{m}{m^2 +
\lambda_{n}^2}\!\gg_{N_f} \label{qcond}
\ee
where $\underline{\lim}$ means the limit $L\to\infty$, followed by $m\to 0$.
The average over gauge configurations is:
\be
\ll \Gamma \gg _{N_f} = Z^{-1} \int d\mu[G]\ e^{-S[G]}\ 
      \Gamma\ \Delta^{N_f}(m|G) \prod_{j>N_f}\Delta (m_j|G), 
     \label{aver}
\ee
where the dependence on $N_f$ is explicit.
This average is conveniently normalized ($\ll\!1\!\gg_{N_f}=1$) and requires a
non-perturbative regularization and renormalization. 
In a similar way, (\ref{VV-AA}) once projected yields\cite{Stern}
\be \label{conduct}
F^2 (N_f) = \underline{\lim}
\frac{1}{L^4} \ll \sum _{k,n} \frac {m}{m^2 + \lambda ^2 _{k}}
\frac {m}{m^2 + \lambda ^2 _{n}} J_{kn} \gg _{N_f},
\ee
with the transition probability:
$J_{kn}=\frac{1}{4}\sum_{\mu} |\int\! dx\
   \phi^\dag_{k}(x)\gamma_\mu \phi_{n}(x)|^2$.
In the chiral limit $m\to 0$, $F^2$ and $\Sigma$ are
essentially dominated by the lowest Dirac e.v.s, because of the
factors $m/(m^2+\lambda^2)$.
The quark condensate is only sensitive to the e.v.s
accumulating like $1/L^4$ in the gluonic average, while
$1/L^2$ is a sufficient speed for an e.v. to contribute to $F^2$.

In $\Delta(m|G)$, infrared and ultraviolet e.v.s can be split
with respect to a cutoff $\Lambda$. If $K$ is the integer such as
$\omega_K=\Lambda$, we have
\be \label{split}
\Delta= m^{|\nu|} \Delta_\mathrm{IR}(m|G) \Delta_\mathrm{UV}(m|G),
\qquad 
\Delta_\mathrm{IR}(m|G) = 
  \prod \limits_{k=1}^{K} \frac{m^2 + \lambda_{k}^{2}[G]}
{m^2 +\omega _{k}^2} < 1.
\ee
$\Delta_\mathrm{UV}$ needs regularization and renormalization. 
However, in (\ref{aver}), 
order parameters dominated by the lowest Dirac e.v.s 
should be essentially sensitive to $\Delta_\mathrm{IR}$, which is bounded by
1 and more and more efficiently suppressed when
$N_f$ goes up. Infrared-dominated order parameters should then
decrease: $\Sigma(N_f+1) < \Sigma(N_f)$ and
$F^2(N_f+1) < F^2(N_f)$. Moreover, since
the quark condensate is more sensitive to the lowest e.v.s than $F^2$, 
its suppression should be stronger.

The dependence of $\Sigma(N_f)$ on the $(N_f+1)$-th quark, called $s$,
is given by the connected part of the correlator:
\be \label{der}
\frac{\partial}{\partial m_s} \Sigma(N_f) = 
\lim_{m\to 0} \int dx \langle 0|\bar u u(x)\ \bar s s(0)|0\rangle^{c} 
   \ \equiv\ \Pi_{Z}(m_s)
\ee
so that $\Sigma(N_f) = \Sigma(N_{f}+1) + \int_{0}^{m_s}d\mu\ \Pi_{Z}(\mu)$.
If the number of flavours lies just below $n_\mathrm{crit}(N_c)$, the large
difference $\Sigma(N_f)-\Sigma(N_{f}+1)$ implies, through $\Pi_{Z}$, a
significant Zweig rule violation in the $0^{++}$ channel 
which is actually observed for $N_f=2-3$. Besides, the
large-$N_c$ limit throws $n_\mathrm{crit}$ to infinity: if
$n_\mathrm{crit}(3)$ lies near 3, the $1/N_c$ expansion should converge very
slowly to the phenomenology in the scalar sector. The study of chiral
phase transitions, far from being academic, could thus deepen our
understanding of actual QCD.

\section{Leutwyler-Smilga sum rules : extensions and applications}

In the $(N_f,N_c)$ plane, SB$\chi$S order parameters could therefore exhibit
very different behaviours for their decrease and their vanishing, leading to
a rich chiral phase structure for QCD. 
The lowest Dirac eigenvalues may shed some light on this problem,
through sum rules first described by Leutwyler and
Smilga\cite{Leutwyler-Smilga}. In a large torus, the lightest
excitations (the pseudoscalar mesons) dominate the partition
function of Euclidean QCD, because heavier states are exponentially suppressed.
Moreover, since periodic boundary conditions are chosen, the effective
Lagrangian density $\mathcal{L}_\mathrm{eff}$
describing the Golstone bosons in a box with periodic boundary conditions
is identical to its
equivalent in an infinite volume\cite{Gass-Leut2} : neither
Lorentz-breaking terms, nor volume-dependent coefficients
arise, once on the torus.

The QCD and $\chi$PT representations of the partition function can thus be
matched for large volumes
\be
\int d\mu[G]\ e^{-S[G]} \det(-iD\!\llap{/}+\tilde M)
 \sim \int d\mu[U] \exp\left[-S_\mathrm{eff}(U,\partial U,Me^{i\theta/N_f})
     \right], \label{matching}
\ee
where $U(x)\in \mathrm{SU}(N_f)$ collects the pseudo-Goldstone bosons,
$\tilde M=(1-\gamma_5)M/2+(1+\gamma_5)M^\dag/2$ is the light quark mass
matrix (with positive real eigenvalues), and $\theta$ the vacuum angle,
conjugate to the winding number $\nu[G]$ in the QCD Lagrangian. 
(\ref{matching}) leads to sum rules for $\langle\sigma_k\rangle_\nu$:
\be
\sigma_k = \sum_{n>0} \frac{1}{\lambda_n^k},
\qquad
\langle \Gamma \rangle_\nu
   = \mathcal{A}_\nu \int_\nu d\mu[G]\ e^{-S[G]} 
        \Big(\prod_{n>0} \lambda_n^2 \Big)^{N_f} \Gamma, \label{gluoav}
\ee
where the functional integral is performed over the
gluonic configurations with a fixed winding number $\nu$, and 
$\mathcal{A}_\nu$ is a normalization constant such that 
$\langle 1 \rangle_\nu=1$.

The sums $\langle\sigma_k\rangle_\nu$ are especially sensitive to the lowest
Dirac eigenvalues, and should therefore reflect SB$\chi$S.
Actually, Leutwyler and Smilga obtained the asymptotic behaviour of such sums
for $L\to\infty$, for instance\cite{Leutwyler-Smilga}
\be
\langle\sigma_2\rangle_\nu
  \to \frac{(V\Sigma)^2}{4[N_f+|\nu|]}, \quad
\langle\sigma_4\rangle_\nu
  \to \frac{(V\Sigma)^4}{16[N_f+|\nu|][(N_f+|\nu|)^2-1]}, \label{lssrst}
\ee
Two questions arise: how would these
limits change in a phase where chiral symmetry is broken but 
$\langle\bar{q}q\rangle$ vanishes ? what are the finite-size 
corrections to these asymptotic results ?

(\ref{lssrst}) were derived by considering only the leading order of
$S_\mathrm{eff}$ in S$\chi$PT. The finite-size corrections to the sum rules
are due to higher orders and should be hard to disentangle even at
intermediate volumes, if their contributions
are small compared to the one due to the quark condensate. On the other hand,
a small $\langle\bar{q}q\rangle$ leave space for large finite-size effects.
G$\chi$PT is designed for this situation\cite{gchipt} and
\emph{e.g.} for $\sigma_4$ at $\nu=0$, 
one ends up with the sum rule\cite{LSSR}:
\bea
\langle \sigma_4\rangle_0
  &=&\frac{V^2}{16N_f(N_f^2-1)}
    \Big\{(V\Sigma^2)^2+(V\Sigma^2)\cdot 4F^2[3Z_S-Z_P-N_fA]\nonumber\\
&& +4F^4[3(Z_S^2+Z_P^2)-2Z_SZ_P+A^2
             -2N_f(Z_S+Z_P)A]\Big\}\!+\!\ldots  \label{lssrgen}
\eea
where $A$, $Z_S$ and $Z_P$ are chiral order parameters arising in the
effective Lagrangian. The dots remind of higher order terms:
(\ref{lssrgen}) neglects contributions that are supposed to remain small for
large volumes.
If $\Sigma=0$, the asymptotic behaviour of
$\langle \sigma_4\rangle_0$ dramatically changes from $V^4$ to $V^2$,
indicating a phase transition\footnote{This result agrees with the
discussion in Sec.~\ref{nfdep} : the $1/L^2$-eigenvalues contribute to $F^2$
(SB$\chi$S), but not to $\Sigma$. Such a situation would take place for
$N_f>n_\mathrm{crit}(N_c)$.}. On the contrary, if the quark
condensate is small but non-vanishing, (\ref{lssrgen}) leads, even at
intermediate volumes, to a large departure from (\ref{lssrst}), and its
shape is related to the detailed pattern of chiral symmetry
breaking ($F^2$, $\Sigma$, $A$, $Z_S$\ldots).

These sum rules might be exploited through lattice simulations, since :

1. the $N_f$-dependence is explicitly displayed in the gluonic average
(\ref{gluoav}), 

2. renormalization group-invariant ratios of the sums $\sigma_k$ can be
studied, 

3. computing the sums requires a set of gluonic configurations ($\nu=0$)
and the resulting lowest Dirac e.v.s, obtained by
diagonalizing the discrete version of
$D\!\llap{/}^2=D^2+\sigma_{\mu\nu}F_{\mu\nu}/2$ to avoid the problematic
doublers from $D\!\llap{/}$.

Once these sums are computed for various (large) volumes,
fits with sum rules similar to (\ref{lssrgen}) could determine chiral order
parameters for the desired point in the $(N_f, N_c)$ plane, until a full
restoration of chiral symmetry.

\section{Conclusion}

Euclidean QCD is a powerful tool to investigate chiral symmetry, since its
spontaneous breaking is reflected by the lowest eigenvalues of the Dirac
operator. This framework provides a qualitative understanding of the
$N_f$-dependence of infrared-dominated order parameters. 
In particular, the observed
Zweig rule violation in the scalar channel could be explained by
a large variation of the quark condensate from 2 to 3 flavours, close to
a chiral phase transition. If forthcoming experiments\cite{daphne}
related to $\pi-\pi$ scattering should soon pin down
$\Sigma(2)$, an extensive study of SB$\chi$S in the $(N_f, N_c)$ plane
remains necessary, and could be undertaken through Leutwyler-Smilga
sum rules joined with lattice computations.
\\
\\
\emph{Acknowledgments}:
It is a pleasure to thank the organizers and the lecturers for the variety and
the quality of the subjects discussed during this school.
Work partly supported by the EEC, TMR-CT98-0169, EURODA$\Phi$NE network.

\end{document}